# Constraints on planetesimal accretion inferred from particle-size distribution in CO chondrites

Gabriel A. Pinto[1,2,*], Yves Marrocchi[1], Alessandro Morbidelli[3], Sébastien Charnoz[4], Maria Eugenia Varela[5], Kevin Soto[6], Rodrigo Martínez[7] & Felipe Olivares[2]

[1]Université de Lorraine, CNRS, CRPG, UMR 7358, Vandœuvre-lès-Nancy, 54501, France
[2] Instituto de Astronomía y Ciencias Planetarias, Universidad de Atacama, Copayapu 485, Copiapó, Chile
[3]Laboratoire Lagrange, UMR7293, Université de Nice Sophia-Antipolis, CNRS, Observatoire de la Côte d'Azur, Boulevard de l'Observatoire, F-06304 Nice Cedex4, France
[4] Université de Paris, Institut de physique du globe de Paris, CNRS, F-75005 Paris, France
[5]Instituto de Ciencias Astronómicas, de la Tierra y el Espacio, ICATE-CONICET, San Juan, Argentina
[6]Facultad de Ciencias, Instituto de Ciencias de la Tierra, Universidad Austral de Chile, Valdivia, Chile
[7]Museo del Meteorito, San Pedro de Atacama, Chile

*Corresponding author: gabriel.pinto@univ-lorraine.fr

**Abstract**

The formation of planetesimals was a key step in the assemblage of planetary bodies, yet many aspects of their formation remain poorly constrained. Notably, the mechanism by which chondrules—sub-millimetric spheroids that dominate primitive meteorites—were incorporated into planetesimals remains poorly understood. Here we classify and analyze particle-size distributions in various CO carbonaceous chondrites found in the Atacama Desert. Our results show that the average circle-equivalent diameters of chondrules define a positive trend with the petrographic grade, which reflects the progressive role of thermal metamorphism within the CO parent body. We show that this relationship could not have been established by thermal metamorphism alone but rather by aerodynamic sorting during accretion. By modeling the self-gravitational contraction of clumps of chondrules, we show that (i) the accretion of the CO parent body(ies) would have generated a gradual change of chondrule size with depth in the parent body, with larger chondrules being more centrally concentrated than smaller ones, and (ii) any subsequent growth by pebble accretion would have been insignificant. These findings give substantial support to the view that planetesimals formed via gravitational collapse.

# 1. Introduction

Planetesimals are solid objects larger than 1 km in diameter that formed by the accumulation of orbiting bodies in the protoplanetary disk and whose internal strengths are dominated by self-gravity; they represent the main building blocks of the planets orbiting the Sun today. Chondrites are fragments of asteroids that were never sufficiently heated to melt their constituent silicates and thus preserve primitive grains of the materials from which they agglomerated, including refractory inclusions and chondrules, cemented together by a complex fine-grained matrix. Refractory inclusions are millimeter- to centimeter-sized particles that represent the oldest dated objects in the solar system (Connelly et al. 2016). Chondrules are (sub-)millimeter-sized igneous spherules that formed by a still elusive high-temperature mechanism linked to either nebular or planetary processes (e.g., Johnson et al. 2015; Marrocchi et al. 2018, 2019). Although chondrules are the main constituents of chondrites and their accretion thus represents a key step in the formation of planetesimals, the mechanism by which they assembled into planetesimals remains poorly constrained.

Recent theoretical advances suggest that planetesimals formed from clumps of small particles (mostly chondrules in the case of chondrites) whose common gravitational attraction outweighed the dispersive action of turbulent diffusion (Klahr & Schreiber 2020, 2021). This process requires that clumps of particles be sufficiently dense and massive and would produce planetesimals of typically ~10 - 100 km in diameter depending on the remaining gas mass in the solar nebula at the time of planetesimal formation. However, the formation of such particle clumps remains debated.

For instance, it was proposed that particles were concentrated into large vortices (Barge & Sommeria 1995) or in regions between small vortices that developed in the disk at the dissipation scale (Cuzzi et al. 2001, 2008). More recently, it was proposed that particle clumps formed due to *streaming instabilities*, hydrodynamic instabilities due to the differential velocities of particles relative to the surrounding gas (Youdin & Goodman 2005; Johansen et al. 2009; Simon & Armitage 2014; Wahlberg Jansson & Johansen 2014, 2017; Johansen et al. 2015; Li et al. 2018, 2019).

For all these scenarios, the effectiveness of streaming instabilities in promoting clump formation depends on particle size, or rather the Stokes number, which here is the ratio between a particle's stopping time due to friction with the gas and the orbital period. For chondrule-sized particles, triggering the gravitational collapse of a pebble cloud in streaming instability (Gerbig et al., 2020) requires particles to be radially concentrated in an annulus (Drążkowska et al. 2016) to achieve a sufficiently large solid/gas ratio (Carrera et al. 2015; Yang et al. 2017). One of the strengths of the gravitational collapse scenarios is that it predicts the formation of binary planetesimals, which are observed in large numbers in the relatively pristine trans-Neptunian belt, and reproduces the observed statistics of their orbital orientations (Nesvorný et al. 2019).

Once planetesimals have formed, they can continue growing by accreting individual particles (if they exceed a critical size of about 1000 km diameter, Ormel & Klahr 2010) as they drift through the disk. This process is known as pebble accretion (Lambrechts & Johansen 2012; Johansen et al. 2015). The initial gravitational contraction of a clump of particles and later pebble accretion should produce characteristic variations of particle size with depth inside the resulting planetesimal. Here we report estimated particle-size distributions within different CO carbonaceous chondrites, chosen because they experienced limited alteration processes after their agglomeration, which could have affected the sizes of their constituent particles. We use our data to model the conditions of planetesimal accretion within the protoplanetary disk and the possible layered structure of the CO parent body(ies).

## 2- Material and method

We surveyed all particles in sections of three CO3 carbonaceous chondrites recovered in the Atacama Desert and provided by the Museo del Meteorito (San Pedro de Atacama, Chile): El Médano 216 (EM 216), El Médano 463 (EM 463) and Los Vientos 123 (LoV 123). We also determined the particle-size distribution in Isna (thick section 3239 from the Muséum national d'Histoire naturelle, Paris, France). Backscattered electron (BSE) mosaics and X-ray compositional maps (Fe, Ni, Al, Mg, Ca, Si, S, Cr) were acquired using (i) a JEOL JSM-6510 scanning electron microscope (SEM) equipped with a Genesis EDX detector and operating with a 3 nA electron beam accelerated at 20 kV (CRPG, Nancy, France) and (ii) a JEOL 6400 SEM operating with a 1 nA electron beam accelerated at 15 kV (Naturhistorisches Museum, Vienna, Austria). The chemical compositions of ferroan olivine grains in FeO-rich chondrules of EM 216 and EM 463 were quantified using a Cameca SX100 electron microprobe at the Service Commun de Microscopies Electroniques et de Microanalyses X (SCMEM, Université de Lorraine, Nancy, France) using a 12 nA focused beam accelerated at 15 kV. LoV 123 was analyzed by wavelength dispersive spectroscopy with an ARL-SEMQ electron microprobe at ICATE (San Juan, Argentina) operating with a 15 nA electron beam accelerated at 15 kV. Natural and synthetic standards were used for both instrument calibrations.

Mosaics of all samples were prepared using the GNU image manipulation program. Particle-size measurements were performed using the Fiji distribution of the ImageJ open-source image processing software (Schindelin et al. 2012). We analyzed all nebular components, including chondrules (types I and II being FeO-poor and -rich, respectively), calcium aluminum-rich inclusions (CAIs, types A and B), amoeboid olivine aggregates (AOAs), and isolated olivine grains (IOGs) over total surface areas of 96 mm$^2$, 27 mm$^2$, 32.4 mm$^2$, and 30.6 mm$^2$ for EM 463, EM 216, LoV 123, and Isna, respectively. The sharpness and continuity of the borders in the X-ray compositional maps was improved in Fiji by first applying mean filter at 1 pixel and then enhancing the image contrast at 0.5%. Each particle was recorded in a mask layer by free-hand tracing (Fig. 1). We did not distinguish between AOAs or type-A or -B CAIs for refractory components in EM 216 and LoV 123. Each particle's diameter ($d$) was calculated assuming that its total area was circular in cross section (i.e., as $d = \sqrt{\text{pixel area}/\pi} \times 2$).

## 3-Results

The $Cr_2O_3$ contents of subhedral FeO-rich olivine crystals were determined for 43, 49, and 62 type-II porphyritic chondrules in EM 216, EM 463, and LoV 123, respectively. Chondrule olivine grains appear heterogeneous in both texture (Fig. 1) and composition ($Fa_{21.3-88.2}$, mean $Fa_{50.2 \pm 10.7}$; Table 3). Ferroan olivines in type-II chondrules of EM 463 contain 0.04–0.57 wt.% $Cr_2O_3$ (average 0.09 ± 0.10 wt.%, 1σ, Table 3); those of EM 216 and LoV 123 contain 0.09 ± 0.10 and 0.30 ± 0.13 wt.% $Cr_2O_3$, respectively (Table 3).

EM 216, EM 463, and LoV 123 show high modal abundances of FeO-poor type-I chondrules (41.59, 33.15, and 46.74%, respectively, Table 1) surrounded by fine-grained Fe-rich matrix (Fig. 1). Isna is dominated by type-II chondrules (33.26%) with occasional type-I chondrules (1.85%, Table 1). The modal abundances of refractory inclusion (CAIs + AOAs) range from 1.59% in Isna to 8.60% in LoV 123 (Table 1). We also observe a large variation in the modal abundances of type-II chondrules, from 5.22% in EM 216 to 33.26% in Isna (Table 1).

Our results show a significant difference between the mean spherical diameters of type-I and type-II chondrules: 92.78 and 162.52 μm, respectively (Table 2). The mean spherical diameters (1σ) of type-I chondrules vary among the different COs: those in Lov 123, EM 216, EM 463, and Isna have average sizes of 71.6 (56.51), 90.91 (65.61), 111.37 (89.44), and 170.81 μm (54.90), respectively (Fig. 2A, Table 2). Similarly, type-II chondrules and refractory inclusions (CAIs + AOAs) show variable average sizes among the different COs (Table 2): in LoV 123, EM 216, EM 463, and Isna, type-II chondrules have mean diameters of 141.64 (112.92), 103.84 (95.26), 164.10 (100.38), and 184.85 μm (80.14), respectively, and refractory inclusions have mean diameters of 62.63 (39.06), 75.32 (52.80), 109.76 (71.24), and 67.11 μm (47.81), respectively. The circularity and the mean aspect ratio of chondrules in our studied CO chondrites is $0.67 \pm 0.15$ and $1.60 \pm 0.45$, respectively.

**Table 1.** Modal abundances (the ratio of component pixel area relative to the total pixel area of the chondrite, in %) of refractory components and chondrules in the analyzed CO3 chondrites.

| Component | LoV 123 | EM 216 | EM 463 | Isna |
|---|---|---|---|---|
| Type I chondrule | 46.74 | 41.59 | 33.15 | 1.85 |
| Type II chondrule | 8.64 | 5.22 | 7.75 | 33.26 |
| *All chondrules* | 55.37 | 46.81 | 40.90 | 35.11 |
| CAI A | - | - | 0.58 | 0.66 |
| CAI B | - | - | 0.55 | 0.93 |
| AOA | - | - | 2.37 | - |
| *All refractory inclusions (RI)* | 8.60 | 6.82 | 3.49 | 1.59 |
| RI/chondrules ratio | 0.16 | 0.15 | 0.09 | 0.05 |

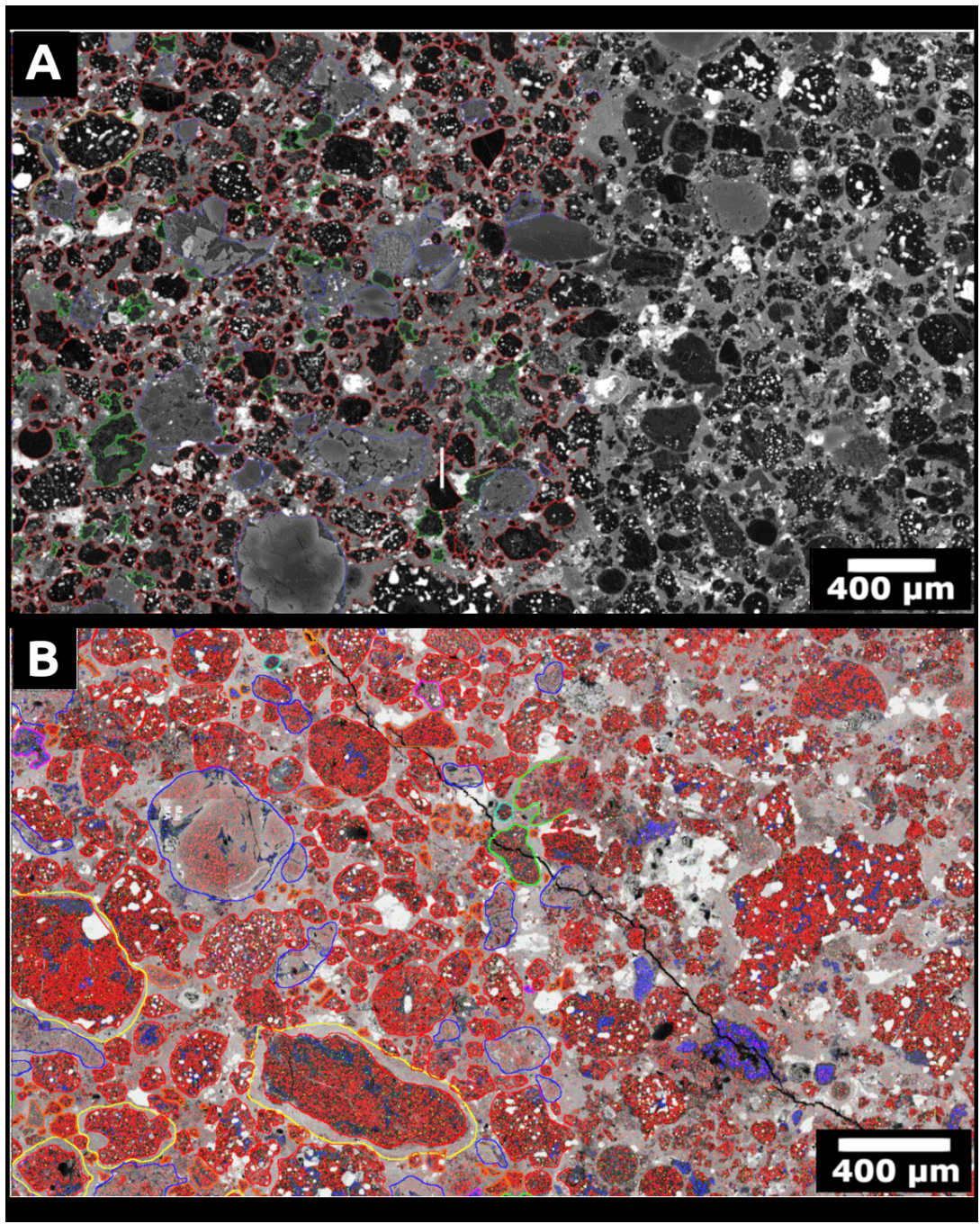

**Figure 1.** Representative examples of particle selection. (A) BSE image of LoV 123. (B) X-ray compositional map of EM 463, with Mg, Al, Ca, and Fe shown in red, blue, green, and white. The colors of particle outlines indicate component type. In (A) type-I chondrules are outlined in red, type-II chondrules in blue, refractory components in green, IOGs in orange, and fine-grained rims in yellow. In (B), outline colors are as in (A), except that refractory components are distinguished between AOAs in green and CAIs in purple.

**Table 2.** The number as well as mean and median diameters (μm) of chondrules and refractory inclusions in each surveyed section.

| Component | LoV 123 | EM 216 | EM 463 | Isna | All CO |
|---|---|---|---|---|---|
| *Type I chondrule* | | | | | |
| n | 2210 | 1140 | 2539 | 25 | 5914 |
| Mean diameter | 71.60 | 90.91 | 111.37 | 170.81 | 92.78 |
| 1σ | 56.51 | 65.61 | 89.44 | 54.90 | 74.92 |
| Median diameter | 55.68 | 71.73 | 88.56 | 158.10 | 72.77 |
| *Type II chondrule* | | | | | |
| n | 109 | 91 | 253 | 318 | 771 |
| Mean diameter | 141.64 | 103.84 | 164.10 | 184.85 | 162.52 |
| 1σ | 112.92 | 95.26 | 100.38 | 80.14 | 97.33 |
| Median diameter | 108.01 | 73.56 | 138.00 | 164.99 | 138.34 |
| *All chondrules* | | | | | |
| Mean diameter | 74.89 | 91.86 | 116.15 | 183.82 | 100.47 |
| 1σ | 62.11 | 68.28 | 89.48 | 71.62 | 80.27 |
| Median diameter | 57.16 | 71.80 | 92.60 | 164.03 | 79.00 |
| *CAIs+AOAs* | | | | | |
| n | 221 | 278 | 252 | 92 | 843 |
| Mean diameter | 62.63 | 75.32 | 109.76 | 67.11 | 81.35 |
| 1σ | 39.06 | 52.80 | 71.24 | 47.81 | 43.08 |
| Median diameter | 53.11 | 61.47 | 88.63 | 49.03 | 64.54 |

## 4. Discussion

### 4.1. Correlation between chondrule-size distribution and degree of metamorphism

The degree of thermal alteration (i.e., petrographic grades) experienced by the carbonaceous chondrites can be estimated based on the mean and standard deviation of $Cr_2O_3$ content of FeO-rich olivines in type-II chondrules (Grossman and Brearley 2005, Table 3). We determined a petrographic grade of LoV 123, EM 216 and EM 462 to be 3.05, 3.2 and 3.2, respectively. The petrographic grade of Isna has been determined to be $3.75 \pm 0.05$ based on the $Cr_2O_3$ content of Fe-rich olivines (Rubin & Li 2019) and detailled petrographic and mineralogical studies of AOAs (Chizmadia et al. 2002). Incorporating recent literature data for CO chondrites (Ebel et al. 2016), average chondrule diameters define a positive relationship with the petrographic grade of their parent chondrite (Fig. 2). Such a correlation was first noticed by Rubin (1989), although they reported a dissimilar relationship (Fig. 2A), likely due to their different estimation method and limited number of analyzed particles (< 1000). Because we used a similar method as Ebel et al. (2016), we here compare our results to their dataset.

The deformation of chondrules (Fig. 1) could have either occurred during the evolution of the protoplanetary disk or within chondritic parent bodies. As lobate chondrules are commonly observed in carbonaceous chondrites that experienced minimal secondary deformation (Jacquet 2021) and CO carbonaceous chondrites show low impact-generated metamorphism transformation (Scott et al., 1992), we thus favor a nebular origin for the non-spherical chondrules.

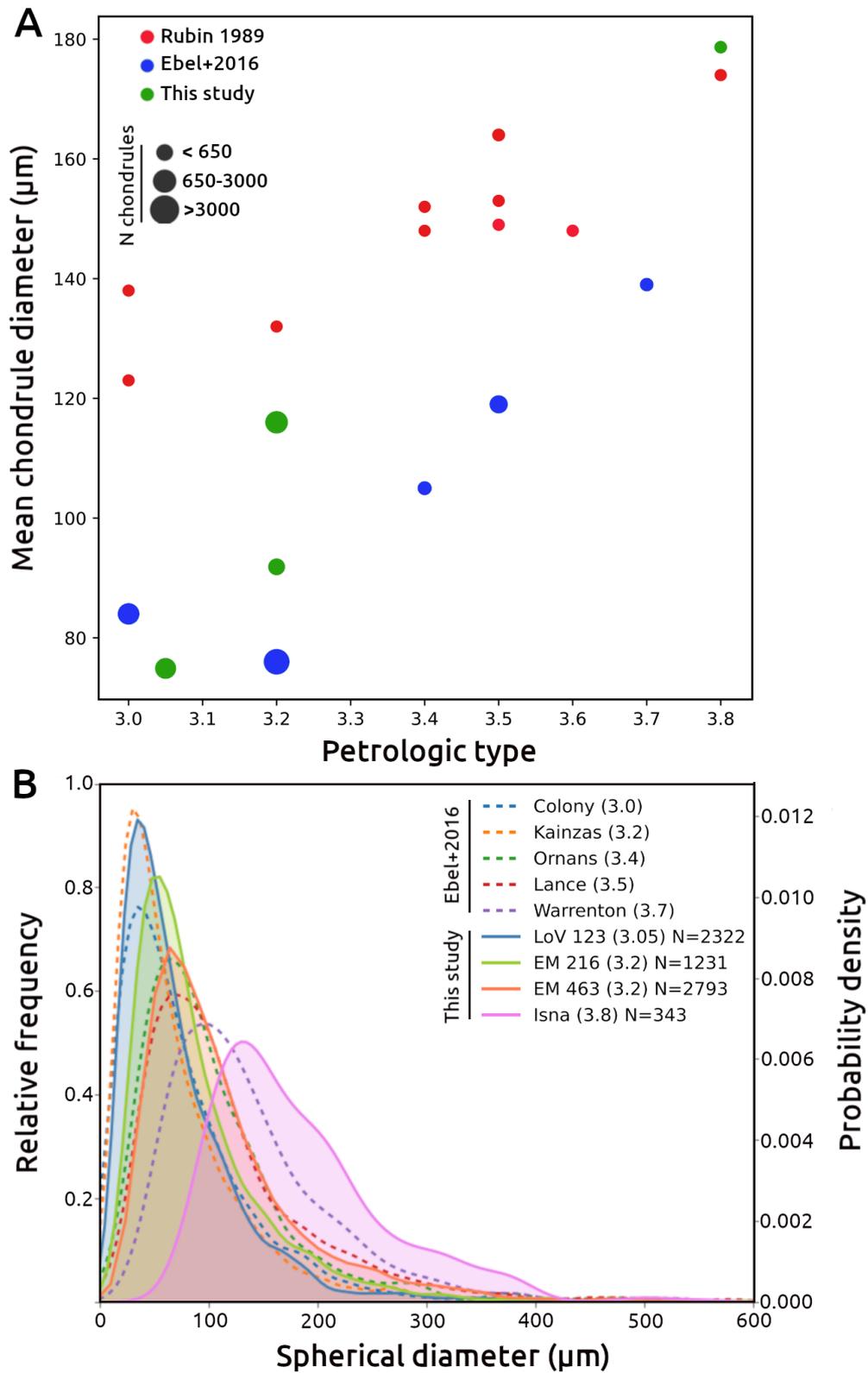

**Figure 2.** (A) Mean chondrule diameter *vs.* petrologic type of CO chondrites (data from Rubin 1989, Ebel et al. 2016, and this study). (B) Probability density function of chondrule diameters in CO chondrites. Solid lines, this study (N, number of analyzed chondrules); dashed lines, Ebel et al. (2016).

The observed correlation between chondrule diameter and petrographic grade in CO chondrites could result from (i) the conditions of planetesimal accretion (Scott & Jones 1990) and/or (ii) post-accretion thermal metamorphism processes resulting from impacts and/or $^{26}$Al decay (Doyle et al. 2015; Vacher et al. 2018; Amsellem et al. 2020; Turner et al. 2021). Although thermal metamorphism could result in mineral coarsening (Huss et al. 2006), this would mainly affect Fe-Ni metal beads and sulfides and would require temperatures >800 °C, significantly hotter than those estimated for CO chondrites (i.e., 300–600 °C; Jones & Rubie 1991; Keil 2000; Bonal et al. 2007). This is consistent with the fact that CV chondrules were only affected by Fe-Mg diffusion without any significant size increase, despite having undergone thermal metamorphism at temperatures higher than in CO chondrites (i.e., ≥600 °C; Ganino & Libourel 2017). Furthermore, type-I chondrule boundaries are well defined in CO chondrites, even in the most metamorphosed sample investigated here (i.e., CO3.8 Isna). Taken together, these lines of evidence indicate that (i) the size characteristics of chondrules result from their formation processes during the evolution of the disk and (ii) aerodynamic sorting played a key role during the accretion of the CO parent body(ies).

Rubin (1989) suggested that the relationship between chondrule size and the degree of metamorphism is related to monotonic changes in the aggregation of materials in the nebular CO reservoir. In this framework, larger chondrules would have been more centrally concentrated in the CO parent body(ies) than smaller chondrules, which could be explained by either simultaneous or sequential accretion of the two chondrule populations (Scott & Jones 1990). Larger chondrules located closer to the center of the CO parent body(ies) would have experienced more protracted thermal metamorphism than smaller chondrules closer to the surface, where heat generated by $^{26}$Al decay was more readily evacuated. Aerodynamic sorting during accretion could thus have produce the co-variation of mean chondrule size with both depth and subsequently with metamorphic grade, as observed for CO chondrites. As CO refractory inclusions are smaller than chondrules (Table 2), this process would also have generated positive relationships between CO metamorphic grade and their (i) bulk oxygen isotopic compositions and (ii) refractory inclusion/chondrule (RI/C) ratios. Interestingly, the former has been reported in previous studies (Clayton & Mayeda 1999; Greenwood & Franchi 2004), and our data confirm that COs with lower metamorphic grades show higher RI/C ratios than more metamorphosed ones (Table 2). These results confirm that aerodynamic sorting during accretion concentrated larger chondrules toward the center of the CO parent body(ies), but smaller chondrules and CAIs were less concentrated at the core. Based on this conclusion, in the following section we model and discuss which accretion process best matches these peculiar features of CO chondrites.

**4.2. Size-sorting during planetesimal formation**

In this section we attempt to provide a qualitative explanation of the interpretation discussed above, in which larger chondrules are more abundant than smaller chondrules at greater depths within the parent planetesimal, and *vice versa*. We first consider the case of planetesimal formation due to the self-gravitational contraction of a clump of particles. Before they collapse into each other, the radial density distribution of particles in the group, $\rho(r)$, is dictated by the equilibrium between gravity and turbulent diffusion of the gas within the group, resulting in (Klahr & Schreiber 2020a, 2020b):

$$\rho(r) = \rho(0) \exp[-r^2/(2l_c^2)], \qquad (1)$$

where

$$l_c = 1/3 \, (\delta/\mathrm{St})^{1/2} H, \qquad (2)$$

where St is the particle's Stokes number, $r$ is the distance to the center, $H$ is the pressure-scaled height of the gas in the disk, and $\delta$ is the non-dimensional coefficient relating the turbulent diffusion coefficient $D$ to the disk's scale height and orbital frequency $\Omega$:

$$D = \delta H^2 \Omega. \tag{3}$$

For particles smaller than the mean free path of gas molecules, St is proportional to particle size (i.e., the Epstein regime). In the Stokes regime, St is proportional to the square of particle size, although this case is rarely considered (nor would it change the considerations below). If there are particles of multiple sizes in the clump, even if the gravitational potential is set by one dominant particle size, each will follow a radial distribution like (1), with its own value of $l_c$. Thus, combining Equations (1) and (2), smaller particles have a more distended radial distribution in the clump, whereas larger particles are more concentrated toward the center. This is because, for larger and smaller particle species 1 and 2, respectively, if $St_1 > St_2$, then $l_{c1} < l_{c2}$.

We now consider the settling of particles towards the center of the clump, forming a solid planetesimal. Particles are accelerated towards the center, but undergo more and more friction as their sedimentation rate increases; the two forces cancel when a particle attains its terminal velocity ($v_t$). Particles accelerated in a gas medium rapidly attain the so-called terminal velocity ($v_t$). The value of the terminal velocity increases approaching the center of gravitational attraction, but the slowest velocity defines the time that the pebble needs to reach the center, and that velocity is set by initial terminal velocity v_t. Thus, in the following we assume, without introducing much error and to simplify the final formulae, that $v_t$ is constant (but is particle-size dependent) during the whole contraction of the clump.

The planetesimal grows over time as more and more particles reach its surface during the sedimentation process. Thus, we can use the accretion timescale $T$ as a proxy for planetesimal radius (not necessarily a linear relationship). In a time interval $dT$ at time $T$, the planetesimal receives the particles that were originally in the clump at a distance between $r$ and $r + dr$ where $r = v_t T$ and $dr = v_t dT$ (here is where the assumption of constant $v_t$ is handy). At time $T$ all particles from specie $i$ ($i=1$ or 2) come from a spherical shell with radius $r_i = V_{ti}T$ and thickness $dr = v_{ti}dT$, so the total mass accumulated in planetesimals between time $T$ and $T + dT$ is $dM_i(T) = 4\pi r_i^2 \rho_i(r_i) dr_i = 4\pi v_{ti}^3 T^2 \rho_i(r_i) dT$. So, the mass ratio of the particles with index 1 and 2 accumulated in the time interval $T$ to $T + dT$ is:

$$\rho_1/\rho_2(T) = (v_{t1}/v_{t2})^3 \cdot [\rho_1(v_{t1}T)/\rho_2(v_{t2}T)], \qquad (4)$$

where the first term $(v_{t1}/v_{t2})^3$ comes from the ratio of the volumes occupied at time 0 (i.e., at the beginning of the contraction of the clump) by the particles that sediment onto the planetesimal surface between $T$ and $T + dT$. Let's now define $x = v_{t1}T/(2\, l_{c1}^2)^{1/2}$. Recall that $v_{t1}/v_{t2} = St_1/St_2$ and $l_{c1}/l_{c2} = (St_2/St_1)^{1/2}$. Thus, $(v_{t2}T)^2/(2l_{c2}^2) = (St_2/St_1)^3 x^2$. By applying the definition (1) into (4) and substituting for x, the mass ratio of the two population of particles landing on the planetesimal at time $T = x(2l_{c1}^2)^{1/2}/v_{t1}$ is:

$$\rho_1/\rho_2(x) = \rho_1(0)/\rho_2(0) \cdot (St_1/St_2)^3 \cdot \exp[-x^2]/\exp[-(St_2/St_1)^3 x^2]. \qquad (5)$$

As an example, Figure 3 shows this function for $St_2 = 0.6 St_1$, which is appropriate for comparing chondrules with diameters of ~80–130 μm, i.e., in the Epstein regime.

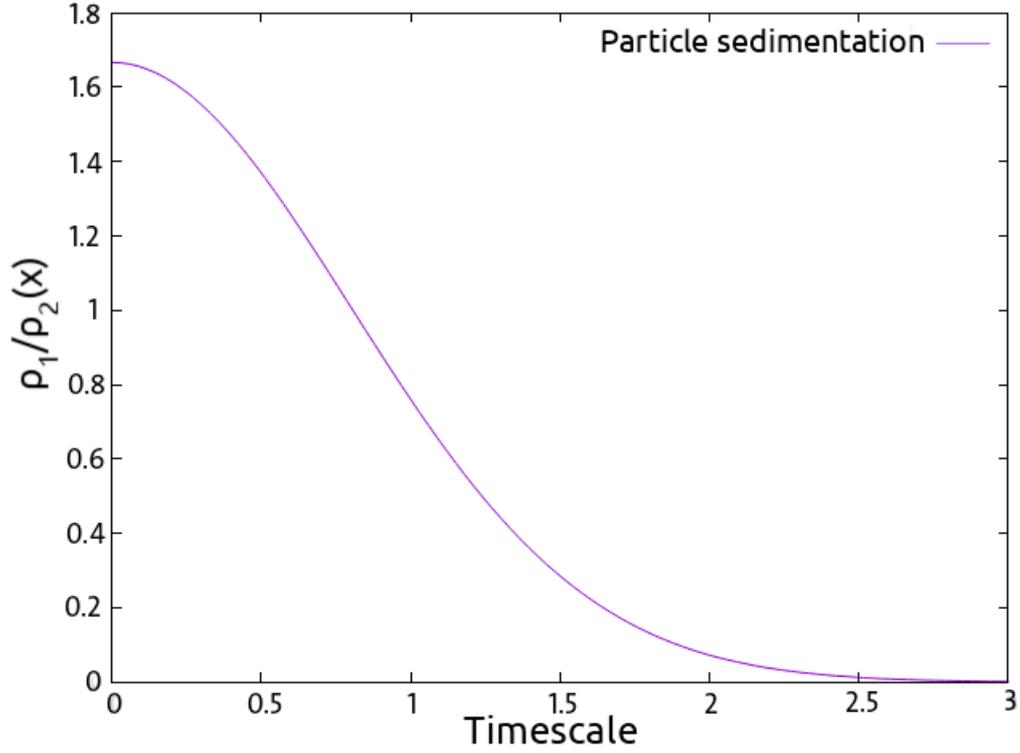

**Figure 3.** Equation (5) normalized to the ratio $\rho_1(0)/\rho_2(0)$ at the center of the particle clump from the beginning of a gravitational contraction leading to the formation of a planetesimal.

Our model shows that the $\rho_1/\rho_2$ decreases monotonically as time progresses (i.e., when the timescale increases). This is in qualitative agreement with our interpretation that average chondrule size increases with depth in the parent body. Of note, chondrules are small compared to pebbles (i.e., 80-130 µm; Table 2) and would have Stokes number of $1 \times 10^{-4}$ at 2 AU for 100 µm particle-size (considering a minimum mass solar nebula). This is smaller than what is typically considered in gravitational collapse models of planetesimal formation (e.g. Yang et al., 2017). To have a more St =1e-3, as more traditionally considered, the gas should have been depleted by a factor of 10. This may well be possible, because chondrites formed at a late time (2-3 My) in the disk's chronology, well after the formation of Jupiter (Kruijer et al., 2017); cavity opening by Jupiter and photo-evaporation may well have reduced the density of gas by one order of magnitude.

Above, we considered the case of a gravitationally contracting planetesimal, but it is well accepted that after their contraction, planetesimals may have grown through pebble accretion, as discussed by Johansen et al. (2015). If a planetesimal accreted particles in the Bondi regime, the Stokes number of the preferentially accreted pebbles increases with the planetesimal's Bondi time, which is proportional to the planetesimal's mass (Lambrechts & Johansen 2012). This predicts that larger chondrules should have been predominant at shallower depths in their parent planetesimal, opposite to our observations. In the case that a larger protoplanet scattered a planetesimal into an orbit of greater inclination, the size of particles accreted by the planetesimal would have suddenly decreased because only small particles were available away from the disk's midplane. However, the mass accreted should have also declined, such that small particles would dominate only in a very thin surface layer (Johansen et al. 2015). We do not observe such a drastic change in chondrule size among the studied meteorites of different petrologic types, but rather a gradual trend. Thus, we conclude that our observations are consistent with the formation of the CO parent body(ies) *via* the contraction of a self-gravitating clump of chondrules of various sizes, and that the subsequent growth of the parent body(ies) due to pebble accretion was insignificant.

## 5. Conclusions

Our particle-size analysis of CO carbonaceous chondrites revealed that the mean spherical diameters of chondrules increase with increasing metamorphic degree. Combining our results with literature data, we show that this relationship was not established during post-accretion thermal metamorphism, but instead records aerodynamic size-sorting of particles during the accretion of the CO parent body(ies). By modeling the self-gravitating contraction of clumps of chondrules, we show that the accretion processes generated a gradual change in chondrule size, with larger chondrules being more centrally concentrated in the parent body(ies) than smaller ones. Our results also suggest that any subsequent planetesimal growth by pebble accretion should have been insignificant. We thus conclude that our observations are consistent with the formation of the CO parent body(ies) via the contraction of a self-gravitating clump of chondrules, supporting the theory that the formation of planetesimals occurred via gravitational collapse.

**Acknowledgments**

We thank Johan Villeneuve, Nicolas Schnuriger, and Laurette Piani for helpful discussion. G.A.P and F.O.E. acknowledge support from FONDECYT project 1201223. G.A.P. was supported by an Eiffel excellence scholarship (grant 968045D). We thank the anonymous review for helpful comments and Frederic Rasio for editorial handling. This is CRPG contribution #2777.

# Appendices

**Table 3.** Chemical compositions of ferroan olivine grains in type-II chondrules of three CO carbonaceous chondrites (EM 463, EM 216, LoV 123).

| CO3 | # | SiO$_2$ | FeO | Al$_2$O$_3$ | CaO | MnO | MgO | Cr$_2$O$_3$ | TiO$_2$ | Total |
|---|---|---|---|---|---|---|---|---|---|---|
| EM 463 | 49 | 38.34 | 27.85 | 0 | 0.09 | 0.31 | 34.58 | 0 | 0 | 101.2 |
| | | 36.05 | 39.18 | 0.03 | 0.11 | 0.28 | 24.89 | 0.15 | 0 | 100.7 |
| | | 37.41 | 30.69 | 0.05 | 0.09 | 0.25 | 31.50 | 0.05 | 0 | 100.0 |
| | | 36.48 | 33.55 | 0.16 | 0.08 | 0.28 | 30.73 | 0.09 | 0 | 101.4 |
| | | 38.51 | 23.92 | 0.03 | 0.01 | 0.31 | 38.32 | 0.01 | 0 | 101.1 |
| | | 38.27 | 26.49 | 0.10 | 0.02 | 0.30 | 36.47 | 0.01 | 0.01 | 101.6 |
| | | 37.83 | 28.32 | 0.10 | 0.14 | 0.34 | 35.69 | 0.14 | 0.01 | 102.6 |
| | | 38.53 | 24.64 | 0.04 | 0.11 | 0.35 | 37.80 | 0 | 0 | 101.5 |
| | | 37.00 | 31.29 | 0.05 | 0.14 | 0.40 | 31.79 | 0.06 | 0.01 | 100.7 |
| | | 39.12 | 23.16 | 0.04 | 0.13 | 0.26 | 39.32 | 0.06 | 0 | 102.1 |
| | | 35.92 | 35.52 | 0 | 0.11 | 0.36 | 28.72 | 0.03 | 0 | 100.7 |
| | | 37.92 | 26.29 | 0.01 | 0.15 | 0.38 | 36.24 | 0.08 | 0 | 101.1 |
| | | 37.91 | 30.51 | 0.06 | 0.09 | 0.17 | 32.96 | 0.08 | 0.01 | 101.8 |
| | | 36.94 | 34.02 | 0.09 | 0.02 | 0.38 | 29.42 | 0.18 | 0 | 101.0 |
| | | 39.29 | 23.36 | 0 | 0.18 | 0.25 | 38.18 | 0.19 | 0.04 | 101.5 |
| | | 40.21 | 17.47 | 0.10 | 0.10 | 0.22 | 42.78 | 0.11 | 0 | 101.0 |
| | | 37.97 | 24.12 | 0.23 | 0.30 | 0.26 | 36.13 | 0.57 | 0.01 | 99.6 |
| | | 38.61 | 26.09 | 0.02 | 0.12 | 0.37 | 35.62 | 0.08 | 0.01 | 100.9 |
| | | 38.40 | 24.24 | 0.09 | 0.09 | 0.16 | 36.61 | 0.08 | 0 | 99.7 |
| | | 38.40 | 26.45 | 0.03 | 0 | 0.22 | 35.04 | 0.05 | 0 | 100.2 |
| | | 37.89 | 30.34 | 0.04 | 0.09 | 0.38 | 33.23 | 0.09 | 0 | 102.1 |
| | | 38.34 | 25.83 | 0 | 0.04 | 0.20 | 35.77 | 0 | 0.02 | 100.2 |
| | | 39.23 | 21.67 | 0.01 | 0.03 | 0.27 | 39.77 | 0.12 | 0 | 101.1 |
| | | 38.30 | 28.82 | 0.08 | 0.13 | 0.43 | 33.43 | 0.16 | 0 | 101.4 |
| | | 37.41 | 31.61 | 0 | 0.18 | 0.31 | 31.75 | 0.04 | 0.03 | 101.3 |
| | | 38.65 | 23.64 | 0.03 | 0.02 | 0.12 | 38.09 | 0.12 | 0 | 100.6 |
| | | 38.63 | 24.94 | 0.10 | 0.06 | 0.16 | 37.24 | 0.19 | 0.01 | 101.3 |
| | | 38.24 | 27.19 | 0.08 | 0 | 0.30 | 35.15 | 0.38 | 0 | 101.3 |
| | | 38.28 | 27.97 | 0.10 | 0.01 | 0.26 | 33.84 | 0.05 | 0 | 100.5 |
| | | 40.41 | 17.10 | 0.03 | 0.07 | 0.22 | 44.05 | 0.03 | 0.03 | 101.9 |
| | | 37.81 | 29.89 | 0.06 | 0.17 | 0.28 | 32.57 | 0.10 | 0 | 100.9 |
| | | 36.27 | 37.11 | 0.02 | 0.10 | 0.48 | 27.79 | 0.07 | 0 | 101.8 |
| | | 36.66 | 33.88 | 0.07 | 0.14 | 0.24 | 30.13 | 0.06 | 0 | 101.2 |
| | | 37.74 | 21.68 | 0.10 | 0.06 | 0.20 | 40.56 | 0.10 | 0 | 100.4 |
| | | 37.55 | 30.09 | 0.01 | 0.11 | 0.22 | 33.38 | 0.04 | 0 | 101.4 |
| | | 37.26 | 32.46 | 0.08 | 0.21 | 0.44 | 31.28 | 0 | 0.05 | 101.8 |
| | | 36.82 | 34.74 | 0 | 0.08 | 0.53 | 29.35 | 0.05 | 0.04 | 101.6 |
| | | 35.60 | 37.13 | 0.03 | 0.37 | 0.19 | 27.65 | 0.02 | 0 | 100.9 |
| | | 35.32 | 40.69 | 0.02 | 0.27 | 0.46 | 23.44 | 0.07 | 0 | 100.3 |
| | | 36.00 | 38.28 | 0.01 | 0.18 | 0.40 | 26.07 | 0.07 | 0 | 101.0 |
| | | 37.63 | 31.64 | 0.05 | 0.14 | 0.43 | 31.66 | 0.13 | 0 | 101.7 |

|  |  |  |  |  |  |  |  |  |  |
|---|---|---|---|---|---|---|---|---|---|
|  |  | 37.12 | 34.90 | 0.04 | 0.16 | 0.45 | 29.55 | 0.08 | 0.01 | 102.3 |
|  |  | 38.95 | 23.89 | 0.01 | 0.03 | 0.16 | 38.35 | 0.07 | 0 | 101.4 |
|  |  | 36.14 | 37.74 | 0.04 | 0.24 | 0.43 | 26.92 | 0.08 | 0 | 101.6 |
|  |  | 38.19 | 27.40 | 0.03 | 0.14 | 0.32 | 35.56 | 0.15 | 0 | 101.8 |
|  |  | 36.79 | 35.39 | 0.01 | 0.22 | 0.39 | 28.60 | 0.08 | 0 | 101.5 |
|  |  | 38.26 | 28.99 | 0.01 | 0.06 | 0.42 | 34.50 | 0.06 | 0 | 102.3 |
|  |  | 36.28 | 35.30 | 0.02 | 0.24 | 0.36 | 28.41 | 0.06 | 0.01 | 100.7 |
|  |  | 37.04 | 29.66 | 0.03 | 0.18 | 0.35 | 33.41 | 0.06 | 0.02 | 100.7 |
|  | **Mean** | **37.71** | **29.33** | **0.05** | **0.12** | **0.31** | **33.56** | **0.09** | **0.01** |  |
|  | **STD** | **1.13** | **5.57** | **0.05** | **0.08** | **0.10** | **4.57** | **0.10** | **0.01** |  |
| EM 216 | 43 | 37.04 | 35.98 | 0.04 | 0.08 | 0.41 | 28.47 | 0.02 | 0 | 102.0 |
|  |  | 37.47 | 33.21 | 0.03 | 0.14 | 0.42 | 30.91 | 0.11 | 0.04 | 102.3 |
|  |  | 36.13 | 38.93 | 0.09 | 0.28 | 0.39 | 26.55 | 0.10 | 0.01 | 102.5 |
|  |  | 37.49 | 28.17 | 0.10 | 0.22 | 0.33 | 35.20 | 0.08 | 0 | 101.6 |
|  |  | 38.52 | 26.03 | 0.01 | 0.06 | 0.30 | 36.89 | 0.11 | 0 | 101.9 |
|  |  | 37.41 | 31.21 | 0.03 | 0.01 | 0.39 | 31.99 | 0.08 | 0 | 101.1 |
|  |  | 35.96 | 39.55 | 0.04 | 0.25 | 0.45 | 24.97 | 0 | 0 | 101.2 |
|  |  | 38.79 | 24.12 | 0.03 | 0.02 | 0.27 | 37.98 | 0.41 | 0 | 101.6 |
|  |  | 37.53 | 33.13 | 0 | 0.15 | 0.27 | 32.12 | 0.09 | 0 | 103.3 |
|  |  | 37.46 | 30.82 | 0.27 | 0.05 | 0.24 | 31.97 | 0.40 | 0 | 101.2 |
|  |  | 36.11 | 39.77 | 0 | 0.14 | 0.22 | 25.86 | 0.03 | 0 | 102.1 |
|  |  | 38.41 | 28.06 | 0 | 0.08 | 0.29 | 34.43 | 0.13 | 0 | 101.4 |
|  |  | 37.11 | 34.77 | 0 | 0.08 | 0.47 | 29.28 | 0.05 | 0 | 101.8 |
|  |  | 38.43 | 25.50 | 0.03 | 0.10 | 0.25 | 36.57 | 0 | 0.02 | 100.9 |
|  |  | 35.91 | 38.17 | 0.08 | 0.19 | 0.48 | 25.92 | 0.19 | 0 | 100.9 |
|  |  | 37.73 | 32.52 | 0.02 | 0.02 | 0.22 | 31.88 | 0.02 | 0 | 102.4 |
|  |  | 35.51 | 41.54 | 0.06 | 0.34 | 0.42 | 23.33 | 0.12 | 0.03 | 101.3 |
|  |  | 36.01 | 37.16 | 0.03 | 0.15 | 0.46 | 27.34 | 0.06 | 0.02 | 101.2 |
|  |  | 38.80 | 25.71 | 0.02 | 0.11 | 0.47 | 37.31 | 0.04 | 0 | 102.4 |
|  |  | 36.89 | 37.57 | 0.02 | 0.06 | 0.37 | 26.87 | 0.11 | 0 | 101.9 |
|  |  | 38.56 | 26.92 | 0 | 0.00 | 0.13 | 35.91 | 0.07 | 0 | 101.6 |
|  |  | 36.18 | 38.87 | 0 | 0.15 | 0.35 | 26.31 | 0.04 | 0 | 101.9 |
|  |  | 36.91 | 31.87 | 0.08 | 0.17 | 0.24 | 31.28 | 0.16 | 0 | 100.7 |
|  |  | 35.93 | 40.76 | 0.02 | 0.22 | 0.55 | 24.52 | 0.05 | 0.04 | 102.1 |
|  |  | 37.44 | 32.88 | 0.06 | 0.05 | 0.47 | 31.49 | 0.02 | 0.00 | 102.4 |
|  |  | 35.80 | 40.93 | 0.01 | 0.28 | 0.38 | 24.63 | 0.08 | 0 | 102.1 |
|  |  | 36.68 | 37.19 | 0.02 | 0.11 | 0.46 | 26.80 | 0.09 | 0.03 | 101.3 |
|  |  | 37.16 | 34.65 | 0.03 | 0.07 | 0.32 | 30.04 | 0.03 | 0 | 102.3 |
|  |  | 35.96 | 39.53 | 0 | 0.07 | 0.19 | 25.62 | 0.06 | 0 | 101.4 |
|  |  | 36.82 | 36.62 | 0 | 0.12 | 0.32 | 28.42 | 0.06 | 0 | 102.4 |
|  |  | 38.34 | 28.48 | 0.04 | 0.02 | 0.32 | 35.09 | 0.09 | 0.02 | 102.4 |
|  |  | 36.15 | 37.60 | 0.03 | 0.07 | 0.32 | 27.12 | 0 | 0.02 | 101.3 |
|  |  | 37.88 | 30.32 | 0.01 | 0.06 | 0.22 | 32.70 | 0.05 | 0.02 | 101.2 |
|  |  | 38.28 | 27.78 | 0 | 0.05 | 0.28 | 35.41 | 0.17 | 0 | 102.0 |
|  |  | 34.40 | 48.35 | 0.01 | 0.14 | 0.45 | 18.21 | 0.07 | 0 | 101.6 |
|  |  | 37.04 | 34.55 | 0.04 | 0.09 | 0.32 | 30.23 | 0.01 | 0 | 102.3 |

|  |  |  |  |  |  |  |  |  |  |
|---|---|---|---|---|---|---|---|---|---|
|  |  |  | 35.51 | 41.70 | 0 | 0.15 | 0.31 | 24.15 | 0.07 | 0 | 101.9 |
|  |  |  | 36.93 | 36.28 | 0.01 | 0.13 | 0.27 | 28.88 | 0 | 0.02 | 102.5 |
|  |  |  | 36.77 | 35.47 | 0.06 | 0.08 | 0.29 | 29.26 | 0.07 | 0 | 102.0 |
|  |  |  | 36.56 | 35.44 | 0.06 | 0.24 | 0.14 | 28.97 | 0.06 | 0 | 101.5 |
|  |  |  | 34.61 | 41.14 | 0.21 | 0.25 | 0.36 | 23.82 | 0.40 | 0 | 100.8 |
|  |  |  | 37.98 | 29.72 | 0 | 0.10 | 0.22 | 33.24 | 0 | 0 | 101.3 |
|  |  |  | 37.46 | 31.18 | 0.06 | 0.10 | 0.28 | 33.48 | 0.18 | 0.01 | 102.7 |
|  | **Mean** | **36.98** | **34.42** | **0.04** | **0.12** | **0.33** | **29.80** | **0.09** | **0.01** |  |
|  | **STD** | **1.09** | **5.44** | **0.05** | **0.08** | **0.10** | **4.51** | **0.10** | **0.01** |  |
| LoV 123 | 62 | 35.67 | 31.03 | - | - | 0.31 | 32.75 | 0.24 | 0 | 100.0 |
|  |  | 35.54 | 33.42 | - | - | 0.31 | 30.54 | 0.23 | 0 | 100.0 |
|  |  | 35.62 | 32.24 | - | - | 0.31 | 32.20 | 0.27 | 0.04 | 100.7 |
|  |  | 35.69 | 33.58 | - | - | 0.20 | 31.47 | 0.27 | 0 | 101.2 |
|  |  | 33.56 | 42.69 | - | - | 0.43 | 23.09 | 0.27 | 0 | 100.0 |
|  |  | 37.40 | 21.97 | - | - | 0.25 | 40.04 | 0.31 | 0 | 100.0 |
|  |  | 33.61 | 41.06 | - | - | 0.33 | 25.03 | 0.32 | 0.03 | 100.4 |
|  |  | 34.18 | 36.98 | - | - | 0.26 | 28.02 | 0.40 | 0 | 99.8 |
|  |  | 36.65 | 27.88 | - | - | 0.22 | 34.58 | 0.13 | 0.04 | 99.5 |
|  |  | 35.73 | 27.90 | - | - | 0.15 | 36.32 | 0.27 | 0.01 | 100.4 |
|  |  | 35.92 | 31.78 | - | - | 0.32 | 33.20 | 0.22 | 0 | 101.4 |
|  |  | 33.28 | 38.93 | - | - | 0.59 | 25.75 | 0.13 | 0.01 | 98.7 |
|  |  | 33.17 | 42.60 | - | - | 0.34 | 23.95 | 0.15 | 0 | 100.2 |
|  |  | 33.48 | 42.07 | - | - | 0.52 | 24.00 | 0.13 | 0.03 | 100.2 |
|  |  | 33.39 | 41.49 | - | - | 0.45 | 24.46 | 0.36 | 0.01 | 100.2 |
|  |  | 32.97 | 42.15 | - | - | 0.46 | 24.82 | 0.25 | 0 | 100.6 |
|  |  | 37.30 | 26.38 | - | - | 0.30 | 35.92 | 0.22 | 0.05 | 100.2 |
|  |  | 40.68 | 20.79 | - | - | 0.23 | 31.63 | 0.33 | 0.12 | 93.8 |
|  |  | 34.73 | 39.52 | - | - | 0.43 | 24.90 | 0.27 | 0.02 | 99.8 |
|  |  | 34.63 | 39.66 | - | - | 0.44 | 24.74 | 0.32 | 0 | 99.8 |
|  |  | 34.68 | 39.59 | - | - | 0.43 | 24.82 | 0.29 | 0.01 | 99.8 |
|  |  | 33.82 | 43.32 | - | - | 0.49 | 22.21 | 0.50 | 0.02 | 100.3 |
|  |  | 35.19 | 38.42 | - | - | 0.46 | 25.68 | 0.30 | 0 | 100.1 |
|  |  | 35.78 | 29.81 | - | - | 0.26 | 33.91 | 0.32 | 0.03 | 100.1 |
|  |  | 35.00 | 34.25 | - | - | 0.32 | 29.95 | 0.21 | 0.01 | 99.7 |
|  |  | 38.25 | 18.43 | - | - | 0.15 | 42.55 | 0.75 | 0 | 100.1 |
|  |  | 34.12 | 38.47 | - | - | 0.52 | 26.47 | 0.36 | 0.01 | 100.0 |
|  |  | 34.13 | 37.41 | - | - | 0.41 | 27.88 | 0.32 | 0.01 | 100.1 |
|  |  | 34.97 | 40.21 | - | - | 0.53 | 24.48 | 0.27 | 0.04 | 100.5 |
|  |  | 34.17 | 42.01 | - | - | 0.62 | 22.93 | 0.10 | 0 | 99.8 |
|  |  | 32.64 | 48.74 | - | - | 0.53 | 18.20 | 0.32 | 0.02 | 100.4 |
|  |  | 32.33 | 52.22 | - | - | 0.61 | 14.18 | 0.25 | 0.03 | 99.6 |
|  |  | 35.88 | 28.68 | - | - | 0.25 | 35.17 | 0.28 | 0.03 | 100.3 |
|  |  | 35.78 | 30.47 | - | - | 0.39 | 33.34 | 0.31 | 0 | 100.3 |
|  |  | 36.19 | 30.39 | - | - | 0.35 | 32.38 | 0.59 | 0 | 99.9 |
|  |  | 36.64 | 30.31 | - | - | 0.21 | 33.17 | 0.24 | 0.02 | 100.6 |
|  |  | 37.65 | 22.20 | - | - | 0.25 | 39.53 | 0.28 | 0 | 99.9 |

|  |  |  |  |  |  |  |  |  |
|---|---|---|---|---|---|---|---|---|
|  | 37.27 | 27.96 | - | - | 0.39 | 34.91 | 0.21 | 0.07 | 100.8 |
|  | 36.62 | 28.41 | - | - | 0.44 | 34.30 | 0.22 | 0 | 100.0 |
|  | 37.66 | 23.47 | - | - | 0.37 | 38.07 | 0.38 | 0.05 | 100.0 |
|  | 35.49 | 36.77 | - | - | 0.32 | 27.34 | 0.21 | 0.04 | 100.2 |
|  | 35.86 | 33.14 | - | - | 0.25 | 30.56 | 0.51 | 0 | 100.3 |
|  | 36.80 | 25.71 | - | - | 0.25 | 36.44 | 0.55 | 0.01 | 99.7 |
|  | 38.40 | 17.55 | - | - | 0.18 | 43.11 | 0.23 | 0.04 | 99.5 |
|  | 36.00 | 30.58 | - | - | 0.36 | 32.64 | 0.22 | 0 | 99.8 |
|  | 37.61 | 24.04 | - | - | 0.22 | 37.69 | 0.45 | 0 | 100.0 |
|  | 37.93 | 24.38 | - | - | 0.30 | 36.94 | 0.34 | 0 | 99.9 |
|  | 36.08 | 31.88 | - | - | 0.34 | 31.55 | 0.16 | 0.03 | 100.0 |
|  | 33.24 | 49.31 | - | - | 0.53 | 16.72 | 0.18 | 0.02 | 100.0 |
|  | 37.14 | 24.45 | - | - | 0.26 | 37.49 | 0.23 | 0.01 | 99.6 |
|  | 40.74 | 6.99 | - | - | 0.11 | 52.25 | 0.25 | 0.02 | 100.3 |
|  | 36.96 | 28.77 | - | - | 0.37 | 33.99 | 0.19 | 0.01 | 100.3 |
|  | 37.19 | 25.44 | - | - | 0.29 | 37.01 | 0.29 | 0 | 100.2 |
|  | 35.33 | 38.88 | - | - | 0.30 | 25.30 | 0.26 | 0.01 | 100.1 |
|  | 38.64 | 15.59 | - | - | 0.35 | 45.25 | 0.66 | 0.03 | 100.5 |
|  | 38.82 | 17.52 | - | - | 0.15 | 43.18 | 0.19 | 0 | 99.9 |
|  | 39.06 | 15.93 | - | - | 0.26 | 44.54 | 0.39 | 0 | 100.2 |
|  | 38.61 | 20.66 | - | - | 0.15 | 40.72 | 0.23 | 0 | 100.4 |
|  | 35.17 | 35.55 | - | - | 0.32 | 28.56 | 0.18 | 0 | 99.8 |
|  | 34.18 | 42.09 | - | - | 0.36 | 22.67 | 0.32 | 0.01 | 99.6 |
|  | 37.36 | 21.05 | - | - | 0.19 | 40.44 | 0.49 | 0.04 | 99.6 |
|  | 36.10 | 29.09 | - | - | 0.23 | 34.65 | 0.26 | 0.05 | 100.4 |
| *Mean* | *35.88* | *31.84* |  |  | *0.34* | *31.62* | *0.30* | *0.02* |  |
| *STD* | *1.92* | *9.27* |  |  | *0.12* | *7.50* | *0.13* | *0.02* |  |